\documentclass{PoS}
\usepackage{amstext}
\usepackage{url}
\title{R-process nucleosynthesis calculations with complete nuclear
physics input}

\ShortTitle{R-process nucleosynthesis}

\author{\speaker{I. Petermann}\thanks{I. Petermann is supported by the
    Deutsche Forschungsgemeinschaft through contract SFB 634.},
  A. Arcones, A. Keli\'c, K. Langanke, G. Mart\'inez-Pinedo, K.-H. Schmidt\\
  GSI
  Helmholtzzentrum f\"ur Schwerionenforschung, Darmstadt, Germany\\
  Institut f\"ur Kernphysik, TU Darmstadt, Germany\\
        }

\author{W. R. Hix\\
        Physics Division, Oak Ridge National Laboratory, Oak Ridge,
        TN37831-6374, USA\\
}

\author{I. Panov, T. Rauscher, F.-K. Thielemann\\
        Department f\"ur Physik und Astronomie, Universit\"at Basel, Switzerland\\
}

\author{N. Zinner\\
        Department of Physics, Harvard University, Cambridge, MA 02138\\
}

\abstract{The r-process constitutes one of the major challenges in
  nuclear astrophysics. Its astrophysical site has not yet been
  identified but there is observational evidence suggesting that at
  least two possible sites should contribute to the solar system
  abundance of r-process elements and that the r-process responsible
  for the production of elements heavier than $Z=56$ operates quite
  robustly producing always the same relative abundances. From the
  nuclear-physics point of view the r-process requires the knowledge
  of a large number of reaction rates involving exotic nuclei. These
  include neutron capture rates, beta-decays and fission rates, the
  latter for the heavier nuclei produced in the r-process. We have
  developed for the first time a complete database of reaction rates
  that in addition to neutron-capture rates and beta-decay half-lives
  includes all possible reactions that can induce fission
  (neutron-capture, beta-decay and spontaneous fission) and the
  corresponding fission yields.  In addition, we have implemented
  these reaction rates in a fully implicit reaction network.  We have
  performed r-process calculations for the neutrino-driven wind
  scenario to explore whether or not fission can
  contribute to provide a robust r-process pattern.}

\FullConference{10th Symposium on Nuclei in the Cosmos\\
		 July 27 - August 1 2008\\
		 Mackinac Island, Michigan, USA}

\begin{document}

\section{Introduction}
The r-process is a series of rapid neutron-capture reactions and
beta-decays in explosive scenarios with high neutron densities. It is
responsible for the synthesis of at least half of the elements heavier
than Fe.  Its astrophysical site has not yet been identified, but
there is observational evidence suggesting the contribution of at
least two sites, namely core collapse supernovae from stars of
different masses and neutron-star mergers.  Observed abundances of
r-process elements with $Z>56$ in metal-poor stars follow a very
robust pattern, that is consistent with scaled solar abundances\cite{metalpoor}.  Our aim is
to explore whether or not fission cycling can provide an explanation
for the robustness.

\section{Fission in the r-process}
In addition to neutron captures and beta decays a full set of fission
rates \cite{cowan}, supplemented by the corresponding fission yields,
are necessary.  We have considered neutron-induced fission, computed
using the SMOKER code \cite{igpa}, beta-delayed fission and
spontaneous fission.  For all of them fission yields have been
computed using the statistical code ABLA \cite{ABLA}, taking the
neutrons evaporated explicitly into account.  The resulting rates have
been implemented in a fully implicit network, that includes
approximately 7000 nuclei from $n,p$ up to $^{300}$Ds. The set of
differential equations is linearized and solved using the
Newton-Raphson method \cite{hixthi}.  The inclusion of fission
maintains the sparseness of the matrix and allows to use sparse matrix
solvers, in our case the Parallel Sparse Direct Linear Solver
(PARDISO) was used~\cite{pardiso}.

Fig. 1 shows the region where fission takes place during the r-process. 
Once the matter-flow breaks out of the magic neutron number $N=184$ 
the fission barriers decrease drastically, and fission becomes possible.

\begin{figure}[hbt]
\centering
\includegraphics[height=4.5cm]{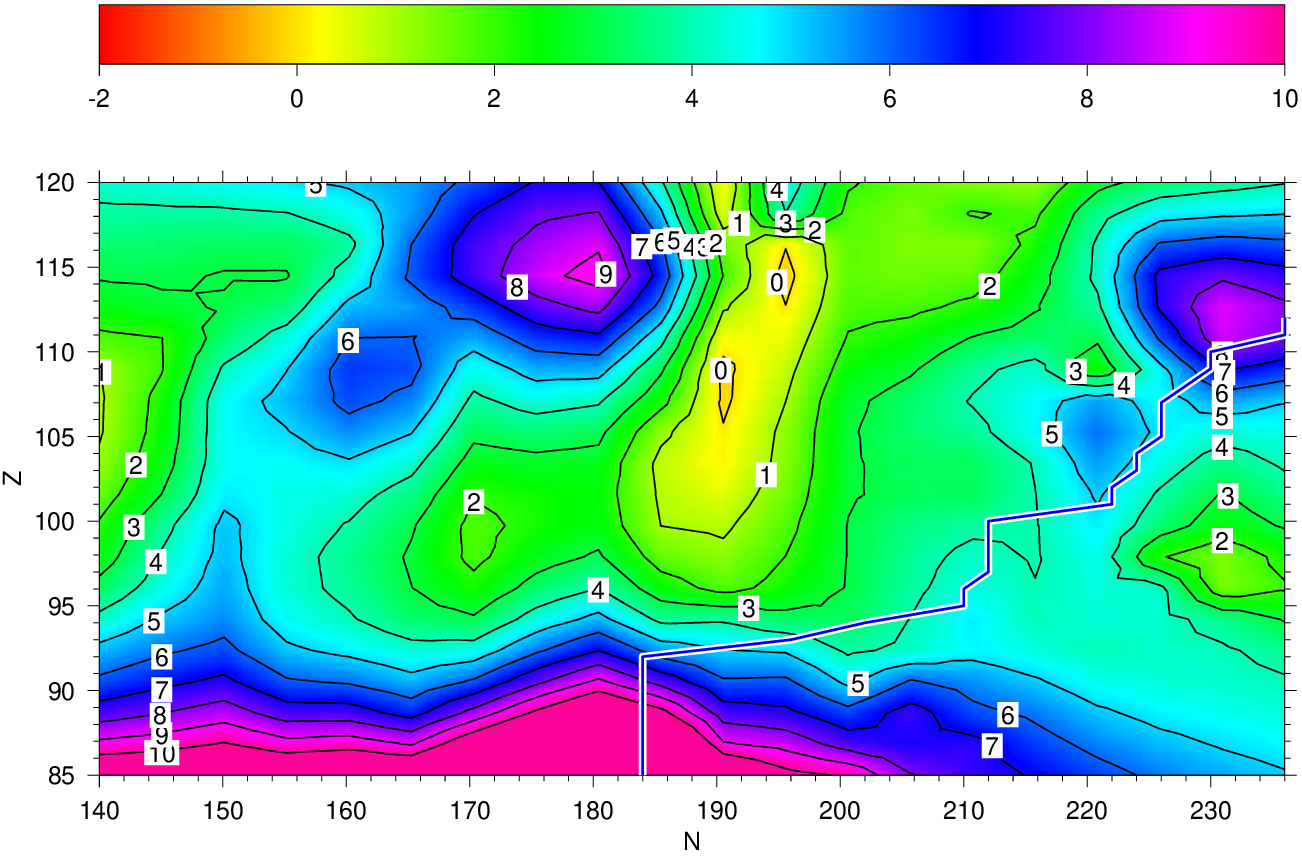}%
\hspace{0.02\linewidth}%
\includegraphics[height=4.5cm]{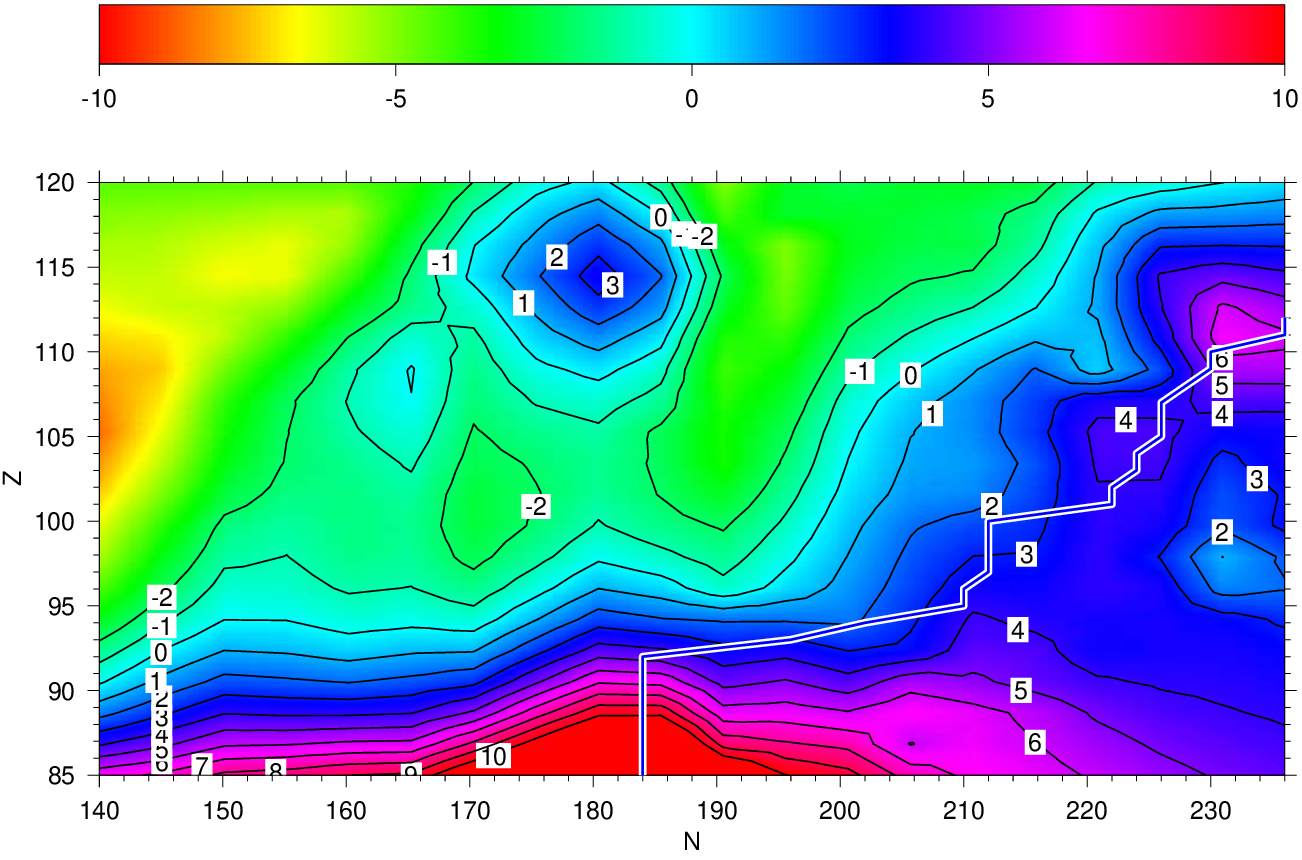}
\caption{Region of the nuclear chart where fission takes place during
  the r-process. In the left panel the contour lines show the
  Myers-Swiatecki \cite{MySwi} fission barrier heights ($B_f$) in MeV.
  The right panel shows contour lines of the quantity $B_f-S_n$, with
  $S_n$ the neutron separation energy taken from the FRDM mass
  model. The unlabeled line shows the location of the
  neutron drip-line using the FRDM mass model.}

\end{figure}

\section{Network calculations}

\begin{figure}[htb]
\centering
\includegraphics[height=5cm]{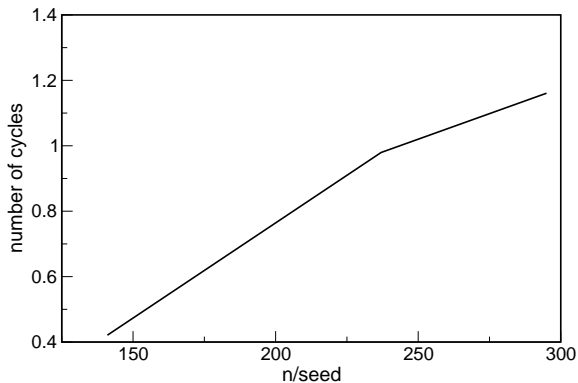}
\caption{Number of fission cycles, for its definition see text, as a
  function of the neutron-to-seed ratio.\label{fcycl}}
\end{figure}

We have performed network calculations using trajectories obtained
from model M15-l1-r1 from the hydrodynamical simulations of
ref.~\cite{almu}. In order to achieve large enough neutron-to-seed
ratios for a successful r-process we have increased the entropy
obtained in these simulations. Our aim is to gain further insight in
the influence that the inclusion of fission has in the observed final
abundances. First, we try to determine what is the neutron-to-seed
ratio necessary such as fission results in a substantial influence in
the final abundances. As an estimate of the amount of fission that has
occured we plot, in figure ~\ref{fcycl}, the number of fission cycles,
defined as $n_{\text{cycl}} =
\log_2(Y_{\text{final}}/Y_{\text{initial}})$, where
$Y_{\text{initial}}$ is the total abundance of nuclei with $A\ge5$
after alpha-rich freeze-out and $Y_{\text{final}}$ is the abundance at
the end of the calculation. At relatively low values of an initial
neutron-to-seed ratio ($\sim 150$), fission already accounts for 40\%
of the final abundance.  This number depends on the mass model used.
In all the calculations shown in this manuscript the FRDM
masses \cite{atommass} were used.

\begin{figure}[htb]
\centering
\includegraphics[height=5.5cm]{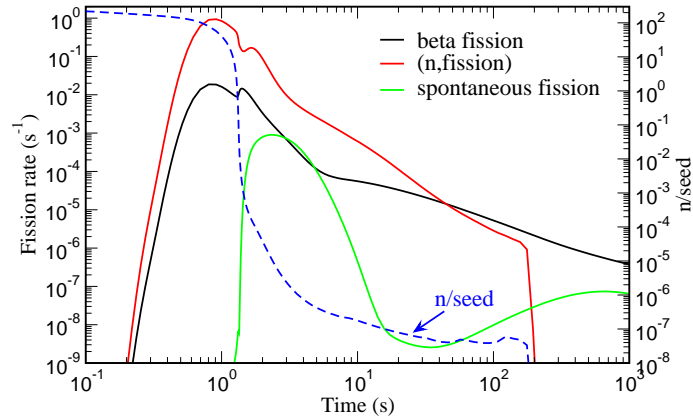}
\caption{Evolution of the fission rate for the different channels for
  a calculation with initial neutron-to-seed ratio 295. Time zero
  corresponds to the beginning of the r-process phase after alpha-rich
  freeze-out. The evolution of the neutron-to-seed ratio is also
  shown (dashed line, right y-axis scale).\label{fig:rfission}} 
\end{figure}

In addition to the total fission rate it is also interesting to
determine what is the dominating fission channel, i.e. neutron induced
fission, beta delayed fission or spontaneous fission.
Fig.~\ref{fig:rfission} shows the evolution of the fission rates for
the different channels. They are computed determining the
instantaneous increase in the total abundance of r-process nuclei,
having $A\ge 5$, due to each fission channel.  Clearly neutron induced
fission is the dominating channel with beta delayed fission having a
slight influence. The end of the r-process can be seen around
1.3 s by the sudden drop in the  neutron-to-seed ratio. 

\begin{figure}[htb]
\centering
\includegraphics[height=5.5cm]{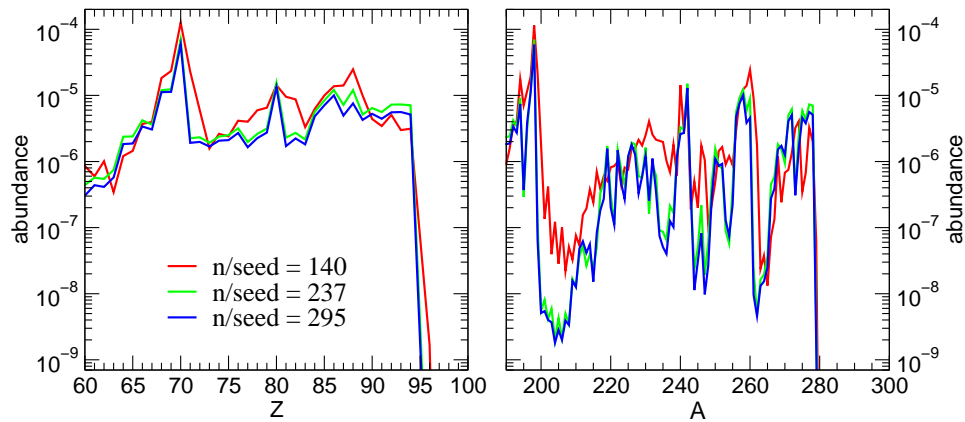}
\caption{Elemental and isotopic abundance distributions for 
$Y_{\text{n}}/Y_{\text{seed}}=140, 237$ and $295$.\label{fig:Fabund}}
\end{figure}

\begin{figure}[htb]
\centering
\includegraphics[height=5.5cm]{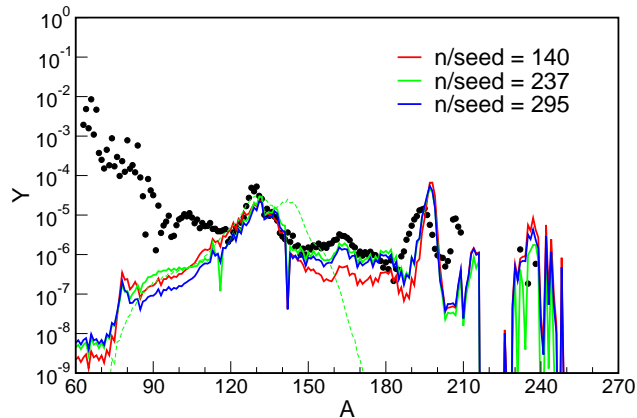}
\caption{R-process abundances resulting from three calculations with
  different neutron-to-seed ratio (solid lines). The symbols show the
  r-process solar distribution. The dashed line shows the integrated
  fission yield distribution, see text for definition, obtained in the
  calculation with n/seed = 295. The other calculations give almost
  identical distributions. \label{fig:solab}}
\end{figure}

The end of the r-process occurs at around 1.3~s and it is
characterized by a sudden drop in the neutron-to-seed ratio. It is
followed by a phase where all heavy fissioning nuclei decay over a
timescale of several hundreds of seconds. During this phase neutron
induced fission still dominates due to the neutrons produced by
fission.  Fig.~\ref{fig:Fabund} shows the elemental and isotopic
abundances for nuclei with $A>190$ for three calculations with
different initial neutron-to-seed ratios, $n/\textrm{seed} = 140, 237$
and 295, at the time when the maximum mean value of $A$ is reached in
each calculation. The heaviest nuclei produced have $Z\sim94$ and
$A\sim280$ with $N=184$. According to our yield calculations, the
fission products of these nuclei are two fragments with $A\sim 132,
141$ and seven neutrons per fission event (see the dashed line in
Fig.\ref{fig:solab}).

The r-process abundances obtained after a time of around 30,000~yr are
shown in Fig.~\ref{fig:solab} for different values of the
neutron-to-seed ratio. Once the number of fission cycles becomes
larger than one a robust r-process abundance distribution is obtained
in the region below the $A \sim 195$ peak. The abundances in this
region are mainly determined by the fission yield distribution. This
distribution is practically the same for the different calculations as
fission occurs in the same region of the nuclear chart (see
fig.~\ref{fig:Fabund}).

The dashed line shows the integrated contribution of fission to the
isobaric abundances, defined as:

\begin{equation}
  Y(A)=\int dt \sum_i \lambda_i^{\text{fis}}(A) Y_i 
\end{equation}
where the sum runs over all the nuclei that can produce a nucleus with
mass $A$ after fissioning with a rate $\lambda_i^{\text{fis}}(A)$. This
distribution shows two peaks. The one at $A\sim 140$ is produced at
early times when nuclei with $A \sim 280$ fission. At these times
neutrons are still present and subsequent neutron captures distribute
the fission yields over higher $A$-values. The peak at $A \sim
130$ is produced at later times when the neutron abundance is rather
small and consequently is much less modified by neutron captures.

\section{Summary}

Using a full set of reaction rates that include all possible fission
channels, supplemented with the corresponding fission yield
distribution, we have explored the influence of fission in the
r-process. Our calculations show that neutron
induced fission is the dominating channel. Similar results were
obtained in ref. \cite{panov} in studies of r-process in neutron star
mergers. The importance of fission has been quantified computing the
number of fission cycles for each r-process calculation. This shows
that already at relatively low initial neutron-to-seed ratios a
noticeable part of the final abundance is due to fission
events. Moreover, calculations with different neutron-to-seed ratios
give very similar final abundance distributions for the mass model
used in this work.

\end{document}